\documentclass[aps,preprint,showpacs,showkeys]{revtex4}
\usepackage{amsfonts}
\usepackage{subfigure}
\usepackage{amsmath}
\usepackage{amssymb}
\usepackage{graphicx}

\begin{document}

\title[ ]{The effect of dipole-dipole interaction on tripartite entanglement
in different cavities}
\author{Salman Khan$^{1}$}
\email{sksafi@comsats.edu.pk}
\author{Munsif Jan$^{2}$}
\affiliation{$^{1}$Department of Physics, COMSATS Institute of Information Technology,
Park Road, Islamabad 45550, Pakistan.}
\affiliation{$^{2}$Department of Physics, Quaid-i-Azam University Islamabad, 45320,
Pakistan}
\date{July 28, 2015}

\begin{abstract}
The effect of dipole-dipole interaction, the initial relative phase and the
coupling strength with the cavity on the dynamics of three two level atoms
in the good and the bad cavity regime are investigated. It is found that the
presence of strong dipole-dipole interaction not only ensures avoiding
entanglement sudden death but also retains entanglement for long time. The
choice of the phase in the initial state is crucial to the operational
regime of the cavity. Under specific conditions, the entanglement can be
frozen in time to its initial values through strong dipole-dipole
interaction. This trait of tripartite entanglement may prove helpful in
engineering multiparticle entanglement for the practical realization of
quantum technology.
\end{abstract}

\maketitle

\section{Introduction}

Entanglement has been recognized as the first candidate in the list of
non-classical correlations and is considered as a vital resource for quantum
information science \cite{Nielsen,Horodecki1}. It has been widely
investigated and a number of schemes have been presented for its detection,
quantification, applications\ and generation \cite{Bennett,Horodecki2,
Wootters,Turchette,Wang,Aspect}. However, the main problem in utilizing its
benefit in quantum information processing is its survival during the
processing of information. Any principle quantum system used for processing
information interacts with environment whose effect on entanglement between
the components of a composite system leads to a number of undesirable
consequences, such as loss of entanglement and entanglement sudden death
(ESD) \cite{Eberly}. On the other hand, study like \cite{Khan} shows that
under certain conditions the environment can generate or reproduce
entanglement. Although initially pointed out theoretically, most of these
phenomenon have now become laboratory facts \cite{Almeida,Aolita}.

Most of the previous studies on the dynamics of entanglement in the presence
of environment are limited to very simplified situations. For instance, the
behavior of entanglement between two atoms under the assumption of equal
coupling strengths between the atoms and the cavity mode is analyzed in Ref.
\cite{Shi,Nicolosi}. This is not a valid assumption while dealing with an
optical cavity. This issue is tackled in \cite{Natali}, however, the effect
of dipole-dipole interaction is ignored as in \cite{Sabrina}, which was
later on addressed in \cite{Yang}. These and other related studies \cite%
{Bing,Jiang} focus on the behavior of bipartite entanglement of two level
atoms. The dynamics of entanglement in two three level atoms and in three
two level atoms has also been investigated \cite{Zhan, Shan}. However, these
studies also ignore the broad perspective and limit their findings to more
simplified scenario.

In this paper, we investigate the dynamics of tripartite entanglement by
considering more realistic situation. That is, we consider three two level
atoms each influenced by its own individual structured reservoir initially
in the vacuum state. Each atom is coupled to its reservoir through a
different coupling strength. The atoms are\ also coupled to each other by
dipole-dipole interaction with a coupling strength different for each pair.
Comparison in the dynamics of entanglement for pure and mixed initial states
is made. The influence of the relative quantum phase in the initial state on
the dynamics of entanglement is demonstrated. We show that through the sole
control of dipole-dipole interaction ESD can safely be avoided. Moreover, it
is shown that the loss of entanglement can not only be slow down in time but
can also be frozen in time by properly tuning the parameters of the cavity.
We expect that due to the relaxed conditions, the results of our model may
easily be realized in the laboratory.

\section{Theoretical Model}

Consider three two-level atoms coupled to each other through dipole-dipole
interaction where each atom is individually influenced by an independent
zero-temperature bosonic reservoir. Under the rotating wave approximation,
the total Hamiltonian of the composite system can be expressed as the sum of
three independent terms in the following way%
\begin{equation}
H=H_{0}+H_{int}+H_{dd},  \label{E1}
\end{equation}%
where $H_{0}$ represents the sum of Hamiltonians of the isolated atoms plus
environments, $H_{int}$ is the atom-environment interaction Hamiltonian and $%
H_{dd}$ represents the dipole-dipole interaction between the atoms. These
can, respectively, be expressed as follows%
\begin{align}
H_{0}& =\omega _{1}\sigma _{+}^{(1)}\sigma _{-}^{(1)}+\omega _{2}\sigma
_{+}^{(2)}\sigma _{-}^{(2)}+\omega _{3}\sigma _{+}^{(3)}\sigma
_{-}^{(3)}+\sum_{k}\omega _{k}b_{k}^{\dagger }b_{k}, \\
H_{int}& =(\alpha _{1}\sigma _{+}^{(1)}+\alpha _{2}\sigma _{+}^{(2)}+\alpha
_{3}\sigma _{+}^{(3)})\sum_{k}g_{k}b_{k}+H.c, \\
H_{dd}& =K_{1}(\sigma _{+}^{(1)}\sigma _{-}^{(2)}+\sigma _{-}^{(1)}\sigma
_{+}^{(2)})+K_{2}(\sigma _{+}^{(2)}\sigma _{-}^{(3)}+\sigma _{-}^{(2)}\sigma
_{+}^{(3)})  \notag \\
& +K_{3}(\sigma _{+}^{(1)}\sigma _{-}^{(3)}+\sigma _{-}^{(1)}\sigma
_{+}^{(3)}).
\end{align}%
In the preceding equations, $\sigma _{\pm }^{(j)}$ $(j=1,2,3)$ represent the
inversion operators and $\omega _{j}$ the transition frequency of the $j$th
atom, $b_{k}^{\dagger }$ $(b_{k})$ represents the creation (annihilation)
operator of the $k$th mode of the environments with eigenfrequency $\omega
_{k}$. The $\alpha _{j}$'s are dimensionless parameters and are measures of
the coupling strength between the cavity and the corresponding atom. These
parameters vary with the strength of the cavity mode and the relative
position of the corresponding atom \cite{Sabrina}. If $\vec{d}_{j}$
represents the dipole moment of an atom and $\vec{r}_{ij}=\vec{r}_{i}-\vec{r}%
_{j}$ ($i\neq j$), ($i=1,2,3$) stands for the relative position, then the
static constants $K_{j}$ are given by \cite{Yang}%
\begin{equation}
K_{j}=r_{ji}^{-3}(\vec{d}_{j}\cdot \vec{d}_{j}-3(\overrightarrow{d}_{j}\cdot
\overrightarrow{r}_{ji})(\overrightarrow{r}_{ji}\cdot \overrightarrow{d}%
_{j})/r_{ji}^{2}).  \label{E2}
\end{equation}%
In Eq. (\ref{E2}) the indices take the values such that when $j=1$, $i=2$; $%
j=2$, $i=3$; and $j=3$, $i=1$.

For an initial state of the form%
\begin{equation}
\left\vert \psi (0)\right\rangle =\left( a\left\vert 100\right\rangle
_{123}+b\left\vert 010\right\rangle _{123}+c\left\vert 001\right\rangle
_{123}\right) \otimes \left\vert \overline{0}\right\rangle ,
\end{equation}%
where $\left\vert 0\right\rangle $, $\left\vert 1\right\rangle $ are the
ground and excited states of an atom, $|a|^{2}+|b|^{2}+|c|^{2}=1$ and $%
\left\vert \overline{0}\right\rangle =\bigotimes_{k}\left\vert
0_{k}\right\rangle $ are the vacuum states of the environment with mode $k$.
In order to investigate the role of relative quantum phase, we will later
replace $b=be^{i\phi }$. Our investigations mainly deal with initial states
correspond to $\phi =\{0,\pi \}$. The time evolution of the composite system
can be written as%
\begin{equation}
\left\vert \psi (t)\right\rangle =c_{11}(t)\left\vert 100\right\rangle
\left\vert 0_{k}\right\rangle +c_{12}(t)\left\vert 010\right\rangle
\left\vert 0_{k}\right\rangle +c_{13}(t)\left\vert 001\right\rangle
\left\vert 0_{k}\right\rangle +\sum_{k}c_{k}(t)\left\vert 000\right\rangle
\left\vert 1_{k}\right\rangle ,
\end{equation}%
where $\left\vert 1_{k}\right\rangle $ represent a single photon in the $k$%
th mode of the reservoir. For mathematical convenience, we consider that the
transition frequency $\omega _{j}$ is same for the three atoms and we denote
it by $\omega _{o}$. The rate equations for the time evolved probability
amplitudes can be obtained from the solution of Schrodinger equation which
take the following form%
\begin{equation}
i\dot{c}_{11}(t)=\alpha _{1}\sum_{k}g_{k}c_{k}(t)e^{-i(\omega _{k}-\omega
_{0})t}+K_{1}c_{12}(t)+K_{3}c_{13}(t),  \label{E3}
\end{equation}%
\begin{equation}
i\dot{c}_{12}(t)=\alpha _{2}\sum_{k}g_{k}c_{k}(t)e^{-i(\omega _{k}-\omega
_{0})t}+K_{2}c_{13}(t)+K_{1}c_{11}(t),  \label{E4}
\end{equation}%
\begin{equation}
i\dot{c}_{13}(t)=\alpha _{3}\sum_{k}g_{k}c_{k}(t)e^{-i(\omega _{k}-\omega
_{0})t}+K_{2}c_{12}(t)+K_{3}c_{11}(t),  \label{E5}
\end{equation}%
\begin{equation}
i\dot{c}_{k}(t)=g_{k}^{\ast }c_{k}(t)e^{i(\omega _{k}-\omega _{0})t}(\alpha
_{1}c_{11}(t)+\alpha _{2}c_{12}(t)+\alpha _{3}c_{13}(t)),  \label{E6}
\end{equation}%
By integrating Eq. (\ref{E6}) and substituting the result into the rest
three rate equations leads to the following form of the probability
amplitudes%
\begin{eqnarray}
\dot{c}_{11}(t) &=&-\int_{0}^{t}dt_{1}G(t-t_{1})\alpha _{1}(\alpha
_{1}c_{11}(t_{1})+\alpha _{2}c_{12}(t_{1})+\alpha _{3}c_{13}(t_{1}))  \notag
\\
&&-iK_{1}c_{12}(t)-iK_{3}c_{13}(t),  \label{E7}
\end{eqnarray}%
\begin{eqnarray}
\dot{c}_{12}(t) &=&-\int_{0}^{t}dt_{1}G(t-t_{1})\alpha _{2}(\alpha
_{1}c_{11}(t_{1})+\alpha _{2}c_{12}(t_{1})+\alpha _{3}c_{13}(t_{1}))  \notag
\\
&&-iK_{2}c_{13}(t)-iK_{1}c_{11}(t),  \label{E8}
\end{eqnarray}%
\begin{eqnarray}
\dot{c}_{13}(t) &=&-\int_{0}^{t}dt_{1}G(t-t_{1})\alpha _{3}(\alpha
_{1}c_{11}(t_{1})+\alpha _{2}c_{12}(t_{1})+\alpha _{3}c_{13}(t_{1}))  \notag
\\
&&-iK_{2}c_{12}(t)-iK_{3}c_{11}(t),  \label{E9}
\end{eqnarray}%
\begin{equation}
c(t)=\sqrt{1-\left\vert c_{11}(t)\right\vert ^{2}-\left\vert
c_{12}(t)\right\vert ^{2}-\left\vert c_{13}(t)\right\vert ^{2}},  \label{E10}
\end{equation}%
where $G(t-t_{1})$ is the correlation function. For the reservoir being
electromagnetic field inside a lossy cavity, the spectral density $J(\omega
_{k})$ is Lorentzian in nature. Under resonant interaction between an atom
and its reservoir, the spectral density can be expressed as follows%
\begin{equation}
J(\omega _{k})=\mid g_{k}\mid ^{2}=\frac{\Re ^{2}\lambda }{\pi \lbrack
(\omega _{k}-\omega _{0})^{2}+\lambda ^{2}]},
\end{equation}%
where the weight factor $\Re $ depends on the vacuum Rabi frequency and $%
\lambda $ stands for the width of the distribution. The relative weight of
these parameters describes the nature of the coupling and hence the
corresponding dynamics of the system. For $\lambda <2\Re $, the coupling is
strong and the dynamics are non-Markovian due to the existence of
oscillatory reversible decay. On the other hand, the dynamics are Markovian
and the coupling is weak when $\lambda >2\Re $ \cite{Dalton} . In terms of
the spectral density, the correlation function becomes%
\begin{equation}
G\left( t-t_{1}\right) =\int_{-\infty }^{\infty }d\omega _{k}J\left( \omega
_{k}\right) e^{-i\left( \omega _{k}-\omega _{0}\right) \left( t-t_{1}\right)
}.
\end{equation}%
With the help of Fourier transform and the residue theorem, this can be
reduced to $G(t-t_{1})=\Re ^{2}e^{-\lambda (t-t_{1})}$, with the inverse of $%
\lambda $ being the reservoir correlation time.

Following the pseudomode approach \cite{Dalton,Garraway} and taking the
Laplace transform, the solutions to the rate equations become%
\begin{equation}
sc_{11}(s)=a-ir_{1}Rb(s)-iK[c_{12}(s)+c_{13}(s)],  \label{E11}
\end{equation}%
\begin{equation}
sc_{12}(s)=b-ir_{2}Rb(s)-iK[c_{13}(s)+c_{11}(s)],  \label{E12}
\end{equation}%
\begin{equation}
sc_{13}(s)=c-ir_{3}Rb(s)-iK[c_{12}(s)+c_{11}(s)],  \label{E13}
\end{equation}%
\begin{equation}
sb(s)-b(0)=-\lambda b(s)-iR[r_{1}c_{11}(s)+r_{2}c_{12}(s)+r_{3}c_{13}(s)],
\label{E14}
\end{equation}%
where the use of pseudomode amplitude%
\begin{equation}
b(t)=-iR\int_{0}^{t}e^{-\lambda
(t-t_{1})}dt_{1}[r_{1}c_{11}(t_{1})+r_{2}c_{12}(t_{1})+r_{3}c_{13}(t_{1})],
\end{equation}%
is made. Also, in the preceding equations, we have introduced the vacuum
Rabi frequency $R=\Re \alpha _{T}$ and the relative coupling strengths $%
r_{j}=\alpha _{j}(\alpha _{T})^{-1}$ with $\alpha _{T}=\sqrt{\alpha
_{1}^{2}+\alpha _{2}^{2}+\alpha _{3}^{2}}$ being the total coupling
strength. To reduce the mathematical rigor and put the equations in more
compact form, we have used a uniform static constant $K_{j}=K$. However, the
effect of $K_{j}$'s being different on the dynamics of entanglement will
also be discussed while depicting it graphically.

The density matrix of the composite system can be constructed from the
probability amplitudes by first applying the inverse Laplace transform to
the preceding equations. Partial tracing over the environment space, leaves
the following reduced density matrix for the system of three atoms%
\begin{eqnarray}
\rho _{G} &=&(1-\left\vert c_{11}(t)\right\vert ^{2}-\left\vert
c_{12}(t)\right\vert ^{2}-\left\vert c_{13}(t)\right\vert ^{2})\left\vert
000\right\rangle \left\langle 000\right\vert +\left\vert
c_{13}(t)\right\vert ^{2}\left\vert 001\right\rangle \left\langle
001\right\vert +  \notag \\
&&\left\vert c_{12}(t)\right\vert ^{2}\left\vert 010\right\rangle
\left\langle 010\right\vert +\left\vert c_{11}(t)\right\vert ^{2}\left\vert
100\right\rangle \left\langle 100\right\vert +(c_{12}(t)c_{13}^{\ast
}(t)\left\vert 010\right\rangle \left\langle 001\right\vert +  \notag \\
&&c_{11}(t)c_{13}^{\ast }(t)\left\vert 100\right\rangle \left\langle
001\right\vert +c_{11}(t)c_{12}^{\ast }(t)\left\vert 100\right\rangle
\left\langle 010\right\vert +H.C).  \label{E15}
\end{eqnarray}%
The explicit form of the relations for the probability amplitudes are given
in the Appendix.

We use tripartite negativity as quantifier for tripartite entanglement. For
a tripartite system, it is defined in terms of three bipartite negativities.
The three bipartite negativities are obtained by dividing the system in a
cyclic manner into three arbitrarily chosen bipartitions. Mathematically, it
is given as follows \cite{Buscemi}%
\begin{equation}
\mathcal{N}^{(3)}(\rho )=\sqrt[3]{\mathcal{N}_{1-23}\mathcal{N}_{2-13}%
\mathcal{N}_{3-12}},  \label{E16}
\end{equation}%
\begin{figure}[h]
\begin{center}
\subfigure[]{\includegraphics[scale=0.8]{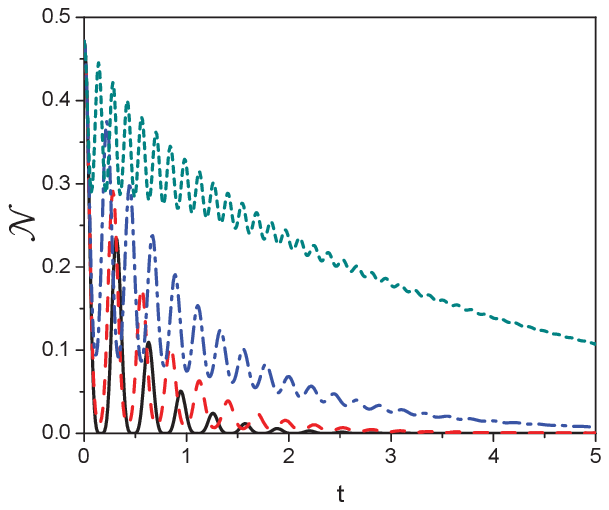}}
\subfigure[]{
\includegraphics[scale=0.8]{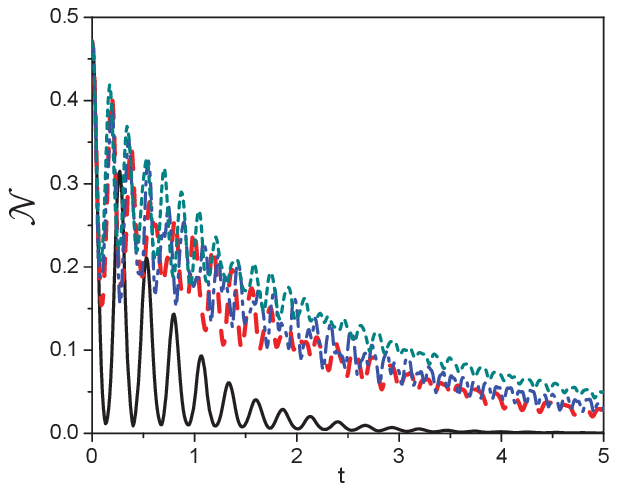}}
\subfigure[]{
\includegraphics[scale=0.8]{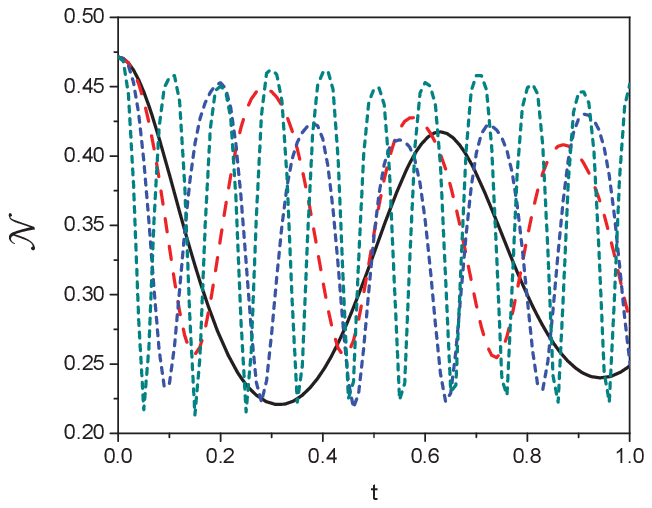}}
\subfigure[]{
\includegraphics[scale=0.8]{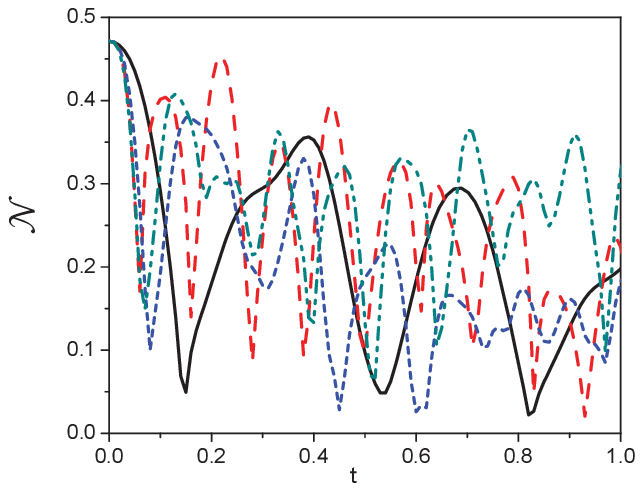}}
\end{center}
\caption{(Color Online) The negativity both for uniform $K$ (1st column) and
non-uniform $K$ (2nd column) is plotted against time $t$ for $\protect\phi %
=0 $ (1st row) and $\protect\phi =\protect\pi $ (2nd row) in the limit of
good cavity. The other parameters in each figure are set to $a=b=c=\frac{1}{%
\protect\sqrt{3}}$, $p=\protect\lambda =1$ and $r_{1}=r_{2}=r_{3}=\frac{1}{%
\protect\sqrt{3}}$ (1st column), $r_{1}=\protect\sqrt{\frac{2}{10}}$, $r_{2}=%
\protect\sqrt{\frac{3}{10}}$, $r_{3}=\protect\sqrt{\frac{5}{10}}$ (2nd
column).}
\label{Fig1}
\end{figure}
where $\mathcal{N}_{i-jk}$ $(k=1,2,3)$ stands for the bipartite negativities
of the three arbitrary bipartitions of the whole system. For each
bipartition, it is given by $\mathcal{N}_{i-jk}=[\sum \lambda _{\mu }(\rho
^{T_{i}})-1]/2$, with $\lambda _{\mu }(\rho ^{T_{i}})$ as the eigenvalues of
the partial transposed density matrix $\rho ^{T_{i}}$\ of the composite
system with respect to subsystem $i$ \cite{Peres,Vidal}.

We focus our analysis of the dynamics of tripartite entanglement to the
following maximally mixed Werner like states%
\begin{equation}
\rho (t)=\frac{1-p}{8}I_{8\times 8}+p\rho _{G},  \label{E17}
\end{equation}%
where the parameter $p$ stands for the purity of the state and $I_{8\times
8} $ is the identity matrix.

The compact relations for the three bipartite negativities become%
\begin{equation}
\mathcal{N}_{1-23}=\frac{1}{4}\left[ -2+\left\vert 1-p\right\vert
+\left\vert \nu _{11}\right\vert +\left\vert \nu _{12}\right\vert
+\left\vert \mu _{1}^{+}\right\vert +\left\vert \mu _{1}^{-}\right\vert %
\right] .  \label{E18}
\end{equation}%
\begin{equation}
\mathcal{N}_{2-13}=\frac{1}{4}\left[ -2+\left\vert 1-p\right\vert
+\left\vert \nu _{21}\right\vert +\left\vert \nu _{22}\right\vert
+\left\vert \mu _{2}^{+}\right\vert +\left\vert \mu _{2}^{-}\right\vert %
\right] .  \label{E19}
\end{equation}%
\begin{equation}
\mathcal{N}_{3-12}=\frac{1}{4}\left[ -2+\left\vert 1-p\right\vert
+\left\vert \nu _{31}\right\vert +\left\vert \nu _{32}\right\vert
+\left\vert \mu _{3}^{+}\right\vert +\left\vert \mu _{3}^{-}\right\vert %
\right] .  \label{E20}
\end{equation}%
where the detail relations for the new parameters are given in the Appendix.
With everything in hand, we are in position to investigate the influence of
different parameters on the tripartite entanglement of the system by using
Eq. (\ref{E16}).

\section{Analysis and discussion}

In figure (\ref{Fig1}), we depict and compare the behavior of negativity
against time for different set of values of $K_{j}$ for two different
initially pure states in the limit of good cavity ($R=10$). The values of
other parameters are given in the caption of the figure. The figures in the
two rows, respectively, correspond to the two initial states different in
phase with $\phi =0$ and $\phi =\pi $. The first column represent the
behavior of negativity for the two states with uniform static constant $K$
for four difference values $K=0,5,10,20$. The second column represents the
situation where each curve corresponds to a different set of values of $%
K_{j} $ whose detail is given below.

\begin{figure}[h]
\begin{center}
\subfigure[]{\includegraphics[scale=0.8]{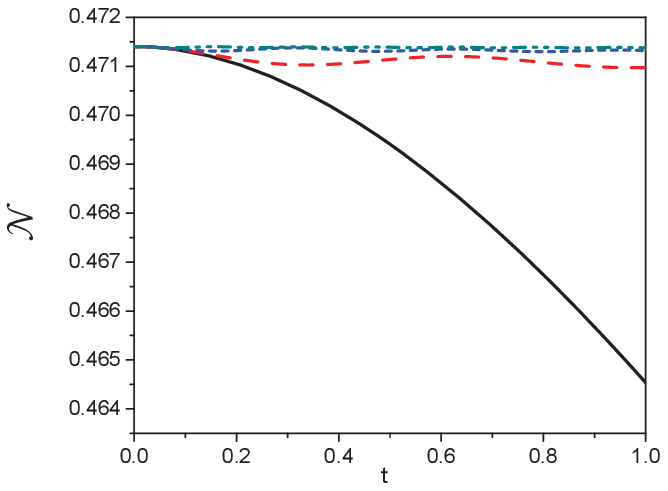}}
\subfigure[]{
\includegraphics[scale=0.8]{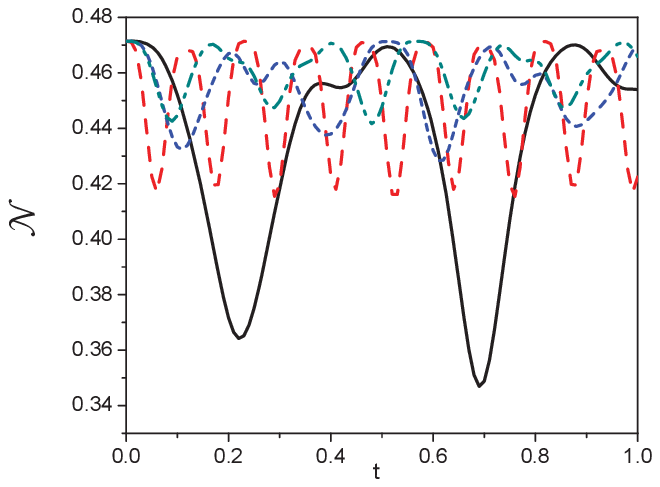}}
\subfigure[]{
\includegraphics[scale=0.8]{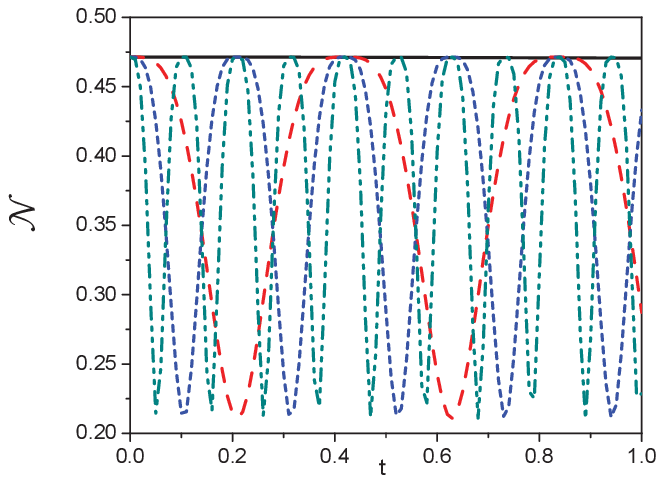}}
\subfigure[]{
\includegraphics[scale=0.8]{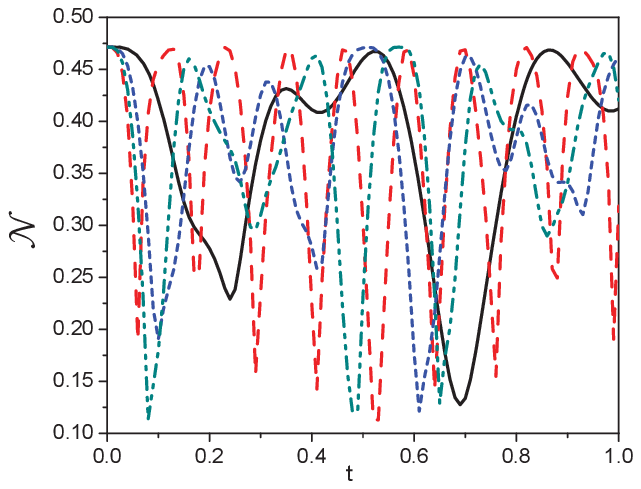}}
\end{center}
\caption{(Color Online) The negativity both for uniform $K$ (1st column) and
non-uniform $K$ (2nd column) is plotted against time $t$ for $\protect\phi %
=0 $ (1st row) and $\protect\phi =\protect\pi $ (2nd row) in the limit of
bad cavity. The other parameters in each figure are set to $a=b=c=\frac{1}{%
\protect\sqrt{3}}$, $p=\protect\lambda =1$ and $r_{1}=r_{2}=r_{3}=\frac{1}{%
\protect\sqrt{3}}$ (1st column), $r_{1}=\protect\sqrt{\frac{2}{10}}$, $r_{2}=%
\protect\sqrt{\frac{3}{10}}$, $r_{3}=\protect\sqrt{\frac{5}{10}}$ (2nd
column).}
\label{Fig2}
\end{figure}

It can be seen (figure(\ref{Fig1}$a$)) that in the absence of dipole-dipole
interaction ($K=0$) the ESD and the revival of entanglement happen
periodically with damped amplitude (solid black curve). This back and forth
process will continue until the entanglement between the atoms is completely
lost. On the other hand, the ESD can be completely avoid for $K\neq 0$. In
this case, with the increase of oscillating frequency in the beginning a
rapid decrease in the amplitude\ of oscillations happens that change it into
a straight line in time with the increasing value of $K$ (from bottom to
top). This behavior of tripartite entanglement is opposite to the bipartite
entanglement in which case the increasing value of $K$ results in the rapid
loss of entanglement \cite{Yang}. However, for $\phi =\pi $ (figure(\ref%
{Fig1}$c$)), the behavior of entanglement is completely different. In order
to see an enlarge view of this different behavior (oscillations), we prefer
to show plots by choosing a relatively short duration of time. In this case,
no ESD occurs for any choice of the values of $K$ and for every $K$ the
entanglement oscillates with damped amplitude about a mean value. For $K=0$
(solid black curve), the oscillation's amplitude damps heavily about the
mean and smoothen to a straight line as the time goes on ($t>5$). For $K\neq
0$, the increasing value of $K$ not only increases the frequency of
oscillations but also increases its amplitude. Moreover, as the value of $K$
increases, it considerably reduces the degree of damping per oscillation.
The red-long dashed curve corresponds to $K=5$, the blue-short dashed curve
is for $K=10$ and the cyan-dashed dotted curve is for $K=20$. For
non-uniform $K$, the ESD can be completely avoided for every set of values
of $K_{j}$'s, however, the entanglement loss may occur at delayed time. In
the case of $\phi =0$ (figure (\ref{Fig1}$b$)), although the amplitude of
oscillation damps with time, however, it may exist for very long and may
effect the periodicity of oscillations depending on the choice of the set of
values of $K_{j}$'s. The set of values of $K_{j}$'s for the curves are,
black ($K_{1}=2$, $K_{2}=5$, $K_{3}=10$), blue ($K_{1}=10$, $K_{2}=15$, $%
K_{3}=20$), red ($K_{1}=2$, $K_{2}=18$, $K_{3}=20$) and cyan ($K_{1}=8$, $%
K_{2}=12$, $K_{3}=18$). Similarly, for $\phi =\pi $, the non-uniformity in $%
K $ not only disturb the periodicity of oscillations but also affect the
amplitude in different ways for every choice of the set of values of $K_{j}$%
's (figure (\ref{Fig1}$d$)). The choices for the set of values of $K_{j}$'s
are the same as in figure (\ref{Fig1}$b$).

\begin{figure}[h]
\begin{center}
\subfigure[]{\includegraphics[scale=0.8]{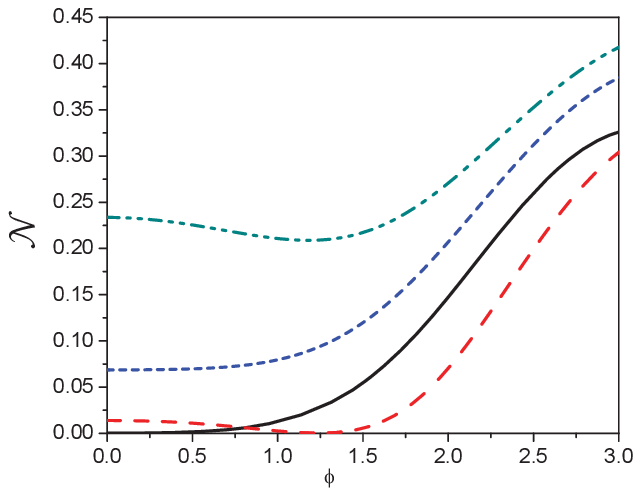}}
\subfigure[]{
\includegraphics[scale=0.8]{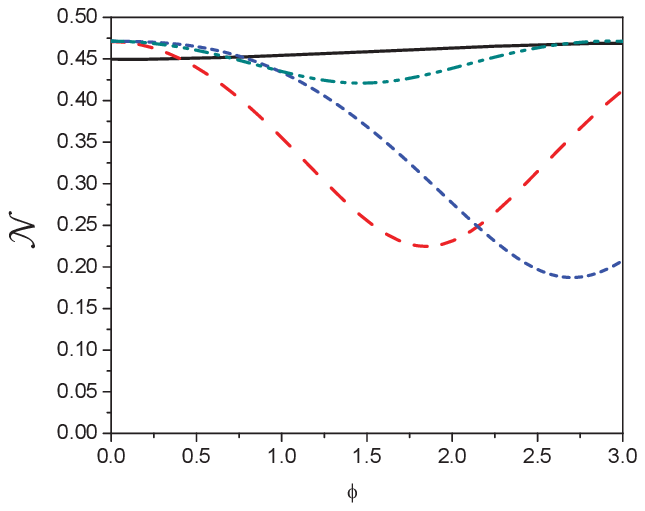}}
\end{center}
\caption{(Color Online) The negativity for uniform $K$ is plotted against
the relative phase $\protect\phi $. The other parameters in each figure are
set to $a=b=c=\frac{1}{\protect\sqrt{3}}$, $p=\protect\lambda =1$ and $%
r_{1}=r_{2}=r_{3}=\frac{1}{\protect\sqrt{3}}.$}
\label{Fig3}
\end{figure}

In figure (\ref{Fig2}) we review figure (\ref{Fig1}) in the limit of bad
cavity ($R=0.1$) with other parameters remain unchanged. It can be seen from
the comparison of the two figures that considerable changes in the behavior
of entanglement in the presence of dipole-dipole interaction take place for
the initial state with $\phi =0$. The entanglement can be frozen to its
initial value in the limit of strong dipole-dipole coupling ($K=20$), the
cyan-short dashed curve in figure (\ref{Fig2}a). For $\phi =\pi $, the
entanglement is static for $K=0$ and unaffected for the rest choices of the
values of $K$ (\ref{Fig2}b). In the case of non-uniform $K_{j}$'s, except
for the large difference in oscillating amplitudes, the behavior of
entanglement for both initial states, qualitatively, becomes identical as
shown in figures (\ref{Fig2}$b$, \ref{Fig2}$d$). The amplitude of
oscillations are considerably damped for the state with $\phi =0$.

Figure (\ref{Fig3}) shows how will the entanglement dynamics be affected by
the strength of dipole-dipole interaction for different choices of the
initial relative phase in the limit of both good (figure (\ref{Fig3}$a$))
and bad (figure (\ref{Fig3}$b$)) cavities. The plots of both the figures are
drawn with uniform $K$. The solid-black curve represents the $K=0$ case, the
red-dashed curve is for $K=5$, the blue-dashed dotted and the cyan-dotted
curve corresponds to $K=10,20$, respectively. In the limit of a good cavity (%
$R=10$), the $\phi =\pi $ is always a good choice while on the contrary the $%
\phi =0$ selection results in more entanglement irrespective of the value of
$K$ in the case of a bad cavity ($R=0.1$). For a good cavity, beyond a
certain limit, the increasing value of $K$ results in more entanglement for
every choice of $\phi $ and get stronger at $\phi =\pi $. On the other hand,
for a bad cavity the best choice is the absence of dipole-dipole interaction.

\begin{figure}[h]
\begin{center}
\subfigure[]{\includegraphics[scale=0.8]{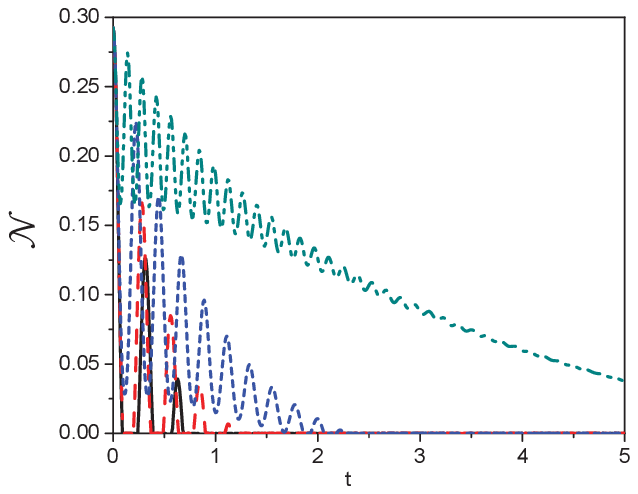}}
\subfigure[]{
\includegraphics[scale=0.8]{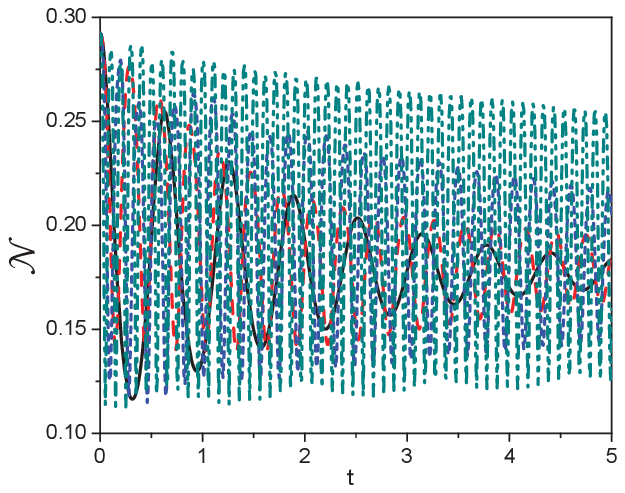}}
\subfigure[]{
\includegraphics[scale=0.8]{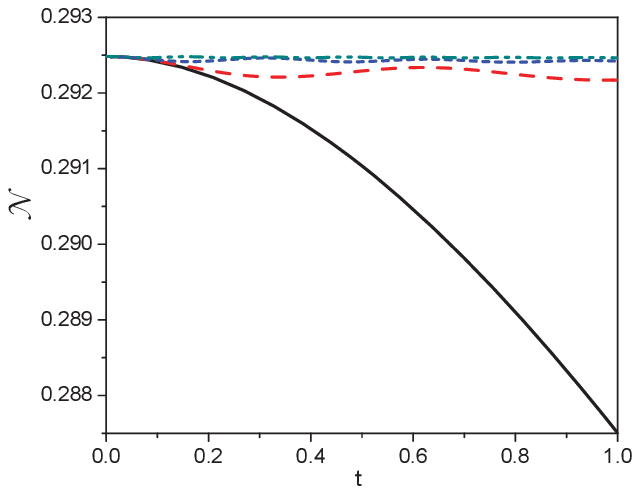}}
\subfigure[]{
\includegraphics[scale=0.8]{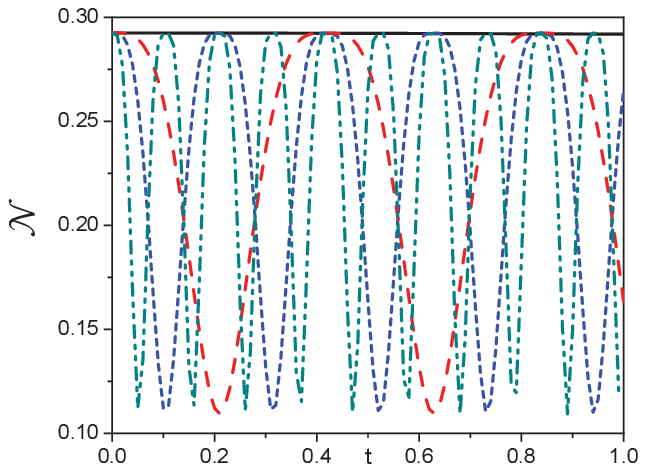}}
\end{center}
\caption{(Color Online) The negativity in the limit of both good (1st row)
and bad (2nd row) cavity for uniform $K$ are plotted for two mixed initial
states $\protect\phi =0$ (1st column) and $\protect\phi =\protect\pi $ (2nd
column) with purity parameter $p=0.7$. The values of other parameters are
set to $a=b=c=\frac{1}{\protect\sqrt{3}}$, $\protect\lambda =1$ and $%
r_{1}=r_{2}=r_{3}=\frac{1}{\protect\sqrt{3}}.$}
\label{Fig4}
\end{figure}

The effect of purity parameter on the time evolution of the negativity is
demonstrated in figure (\ref{Fig4}). The first row shows its behavior in a
good cavity and the second row represents its behavior in the limit of a bad
cavity. The two columns, respectively, correspond to the initial states with
relative phases $\phi =0$ and $\phi =\pi $. The values of all the parameters
are the same as in the first column of figure (\ref{Fig1}) except for $p=0.7$%
. A comparison with the first column of figure (\ref{Fig1}) reveals that for
initial state with $\phi =0$, the overall behavior of negativity is
qualitatively unchanged in the case of strong dipole-dipole interaction (the
cyan-short dashed curve). However, for $K=0$ (solid black curve) and for
weak dipole-dipole interaction, the ESD occurs which is followed by its
revival in a way that in each subsequent turn the amplitude is strongly
damped which results shortly in the complete loss of entanglement. On the
other hand, for the case of $\phi =\pi $ (figure (\ref{Fig4}$b$)), besides
the initial overall damping the rest of the behavior, irrespective of the
values of $K$, is qualitatively unchanged. Similarly, a comparison of the
second row with the first column of figure (\ref{Fig2}) reveals that the
qualitative behavior of pure and mixed initial states in the limit of bad
cavity are identical.

\section{\textbf{Conclusion}}

In this paper, we have investigated the influence of dipole-dipole
interaction, initial relative quantum phase and the coupling strength of the
system with the cavity on the behavior of entanglement between three two
level atoms in the limit of both good and bad cavities. Comparison in the
dynamics of tripartite entanglement between the choices of different initial
states in both good and bad cavities under different sets of the controlling
parameters is made. For a pure initial entangled state with zero relative
phase in a good cavity, it is shown that the presence of strong
dipole-dipole interaction not only avoid ESD but also enhances the retention
of entanglement for long time both for uniform and non-uniform coupling of
the system with the cavity. This behavior of tripartite entanglement is
counterintuitive from the perspective of bipartite entanglement. Most
importantly, it is shown that in the regime of bad cavity, the entanglement
can be frozen to its initial value under strong dipole-dipole interaction.
This behavior of tripartite entanglement is important for the practical
realization of many suggested protocols for quantum information processing.
Moreover, it is found that the choice of the phase in choosing initial state
of the system is crucial to the characteristic of the cavity. We believe
that our results may prove helpful to engineer multiparticle entanglement in
laboratory for the accomplishment of many task in the foreseen quantum
technology.

\textbf{Acknowledgment}

Munsif Jan is thankful to Razmi Fellowship for partially supporting this
work under the Financial Assistance Program for Quaid-i-Azam University
Islamabad, 45320, Pakistan.

\section{Appendix}

The explicit form of the probability amplitudes are given as follows%
\begin{eqnarray}
c_{11}(s) &=&\frac{-be^{i\phi
_{1}}(K_{2}K_{3}+iK_{1}s)-c(K_{1}K_{2}+iK_{3}s)+a(K_{2}^{2}+s^{2})}{%
s^{3}+s(K_{1}^{2}+K_{2}^{2}+K_{3}^{2})-2iK_{1}K_{2}K_{3}}+ \\
&&\frac{%
ib(s)R(-r_{1}K_{2}^{2}+K_{2}(K_{3}r_{2}+K_{1}r_{3})+is(K_{1}r_{2}+K_{3}r_{3}+ir_{1}s))%
}{s^{3}+s(K_{1}^{2}+K_{2}^{2}+K_{3}^{2})-2iK_{1}K_{2}K_{3}},  \notag
\end{eqnarray}%
\begin{eqnarray}
c_{12}(s) &=&\frac{-a(K_{2}K_{3}+iK_{1}s)-c(K_{1}K_{3}+iK_{2}s)+be^{i\phi
_{1}}(K_{3}^{2}+s^{2})}{%
s^{3}+s(K_{1}^{2}+K_{2}^{2}+K_{3}^{2})-2iK_{1}K_{2}K_{3}}+ \\
&&\frac{%
ib(s)R(K_{3}(K_{2}r_{1}-K_{3}r_{2}+K_{1}r_{3})+is(K_{1}r_{1}+K_{2}r_{3})-r_{2}s^{2})%
}{s^{3}+s(K_{1}^{2}+K_{2}^{2}+K_{3}^{2})-2iK_{1}K_{2}K_{3}},  \notag
\end{eqnarray}%
\begin{eqnarray}
c_{13}(s) &=&\frac{-be^{i\phi
_{1}}(K_{1}K_{3}+iK_{2}s)+a(K_{1}K_{2}+iK_{3}s)-c(K_{1}^{2}+s^{2})}{%
s^{3}+s(K_{1}^{2}+K_{2}^{2}+K_{3}^{2})-2iK_{1}K_{2}K_{3}}+ \\
&&\frac{%
b(s)R(iK_{1}(-K_{2}r_{1}-K_{3}r_{2}+K_{1}r_{3})+s(K_{3}r_{1}+K_{2}r_{2})+ir_{3}s^{2})%
}{s^{3}+s(K_{1}^{2}+K_{2}^{2}+K_{3}^{2})-2iK_{1}K_{2}K_{3}}.  \notag
\end{eqnarray}%
The parameters in Eq. (\ref{E18}) are given as follows%
\begin{eqnarray}
\mu _{1}^{\pm } &=&\nu _{13}\pm \sqrt{p(\nu _{14}-\nu _{15})},  \notag \\
\nu _{11} &=&\frac{1-p}{4}+2p\left\vert c_{11}(t)\right\vert ^{2},  \notag \\
\nu _{12} &=&\frac{1-p}{4}+2p\left( 1-\left\vert c_{11}(t)\right\vert
^{2}-\left\vert c(t)\right\vert ^{2}\right) ,  \notag \\
\nu _{13} &=&\frac{1-p}{4}+p\left\vert c(t)\right\vert ^{2},  \notag \\
\nu _{14} &=&\frac{1-p}{4}+4p\left\vert c_{11}(t)\right\vert ^{2}\left(
1-\left\vert c_{11}(t)\right\vert ^{2}\right) ,  \notag \\
\nu _{15} &=&\frac{1-p}{4}+p\left\vert c(t)\right\vert ^{2}\left(
4\left\vert c_{11}(t)\right\vert ^{2}-\left\vert c(t)\right\vert ^{2}\right)
.
\end{eqnarray}%
The parameters in Eq. (\ref{E19}) are given as follows%
\begin{eqnarray}
\mu _{2}^{\pm } &=&\nu _{13}\pm \sqrt{p(\nu _{23}-\nu _{24})},  \notag \\
\nu _{21} &=&\frac{1-p}{4}+2p\left\vert c_{12}(t)\right\vert ^{2},  \notag \\
\nu _{22} &=&\frac{1-p}{4}+2p\left( 1-\left\vert c_{12}(t)\right\vert
^{2}-\left\vert c(t)\right\vert ^{2}\right) ,  \notag \\
\nu _{23} &=&\frac{1-p}{4}+4p\left\vert c_{12}(t)\right\vert ^{2}\left(
1-\left\vert c_{12}(t)\right\vert ^{2}\right) ,  \notag \\
\nu _{24} &=&\frac{1-p}{4}+p\left\vert c(t)\right\vert ^{2}\left(
4\left\vert c_{12}(t)\right\vert ^{2}-\left\vert c(t)\right\vert ^{2}\right)
.
\end{eqnarray}%
The parameters in Eq. (\ref{E20}) are given as follows%
\begin{eqnarray}
\mu _{3}^{\pm } &=&\nu _{13}\pm \sqrt{p(\nu _{33}-\nu _{34})},  \notag \\
\nu _{31} &=&\frac{1-p}{4}+2p\left\vert c_{13}(t)\right\vert ^{2},  \notag \\
\nu _{32} &=&\frac{1-p}{4}+2p\left( 1-\left\vert c_{13}(t)\right\vert
^{2}-\left\vert c(t)\right\vert ^{2}\right) ,  \notag \\
\nu _{33} &=&\frac{1-p}{4}+4p\left\vert c_{13}(t)\right\vert ^{2}\left(
1-\left\vert c_{12}(t)\right\vert ^{2}\right) ,  \notag \\
\nu _{34} &=&\frac{1-p}{4}+p\left\vert c(t)\right\vert ^{2}\left(
4\left\vert c_{13}(t)\right\vert ^{2}-\left\vert c(t)\right\vert ^{2}\right)
.
\end{eqnarray}

\end{document}